
\input harvmac
\input epsf
\def \sign{ \mathop{ \rm sign}\nolimits}
\def \trace{ \mathop{ \rm trace}\nolimits}

\Title{DFTUZ /95/05}
{\vbox{\centerline{Non-diagonal solutions to reflection equations }
	\vskip2pt\centerline{ in $su (n) $ spin chains }}}
\centerline{J. Abad and M. Rios}
\centerline{Departamento de F\'{\i}sica Te\'{o}rica, Facultad de Ciencias,}
\centerline{Universidad de Zaragoza, 50009 Zaragoza, Spain}
\bigskip
\bigskip
\vskip .3in
\centerline{ \tenbf Abstract}
The reflection equations in a $su(3)$ spin chain with open boundary conditions
are
 analyzed. We find non diagonal solutions to the boundary matrices. The
corresponding hamiltonian  is given. The solutions are  generalized to $su(n)$
{}.
\bigskip
\bigskip
\Date{}

\vfill
\eject

The search for integrable models through the Yang-Baxter algebra has deserved
considerable attention in the last years.
For this purpose, quantum group representations give a systematic tool to
obtain integrable systems.
An aspect to take into account is which are the boundary conditions that those
systems must verify
 in order to preserve integrability.
In this line, the  biggest development has been made for systems with periodic
boundary conditions. So, the systems associated to the fundamental
representations
 of the algebras $A_{(r)} $ can be found in ref. \ref\ri{H.J. de Vega, Int. J.
Mod. Phys. A4, (1989) 2371.} and references therein. Other models associated to
the algebra  $A_{(1)} $ in nilpotent representations can be found in
ref. \ref\rii{C. Gomez and G. Sierra, Phys. Lett. B285, (1992) 126.}. Actually
there is a lot of literature about models associated to other types of
algebras.

An other aspect of the problem is to consider systems with open boundary
conditions preserving integrability \ref\riii{V. Pasquier and H. Saleur, Nucl.
Phys. B  330, (1990) 523.},\ref\riv{E. K. Sklyanin, J. Phys. A21, (1988)
2375.}.
The theoretical framework of the problem was set up in ref. \riv\ where
boundary matrices are introduced and the relations they must fulfill in systems
invariant under parity $P$, time reversal $T$ and crossing unitarity are
obtained, this formalism  is extended the refs. \ref\rv{L. Mezincescu and R. I.
Nepomechie, Int. J. Mod. Phys. A 6, (1991) 5281.}%
\nref\rvi{L. Mezincescu and R. I. Nepomechie, Int. J. Mod. Phys. A 7, (1992)
5657.}%
\nref\rvii{L. Mezincescu and R. I. Nepomechie, Jour. of Phys. A 24, (1991)
L7.}%
--\ref\rviii{N. Y. Reshetikhin and M. Semenov-Tian-Shansky, Lett. Math. Phys.
19,
(1990) 133.} to systems symmetric under $PT$ and non-invariant under crossing
 unitarity.  A careful analysis in the framework of the algebraic structures
was carry out in refs. \ref\rvj{P. P. Kulish, R. Sasaki and C. Schwiebert, J.
Math. Phys. 34, (1993) 286.} and
\ref\rvjj{P. P. Kulish and E. K. Sklyanin, J. Phys. A25, (1992) 5963.}.
The general solution for the boundary matrices in the models $XXX$, $XXZ$ and
$XYZ$ can be found in \ref\rix{H. J. de Vega and A Gonz\'{a}lez-Ruiz, Boundary
K-matrices for the $XYZ$, $XXZ$ and $XXX$ spin chains, preprint LPTHE-PAR
93-29 (hep-th/9306089).}--\ref\rjjj{T. Inami and H. Konno, J. Phys. A 27,1994)
L913.}. Also diagonal
solutions for the models $A_{(r)}$, $r\geq 2$ in  their fundamental
representation
are obtained in
\ref\rx{H. J. de Vega and A Gonz\'{a}lez-Ruiz, Jour. of Phys. A 26, (1993)
L519.} .

In this paper, we show non-diagonal solutions for the boundary matrices in
models
based on the algebras $A_{(r)}$, $r\geq 2$. To find them, we must solve a set
of equations whose number increases exponentially with $r$, although the number
of resulting incompatibilities that appears, which leading to simplifications,
increases too.

As is well known, to find integrable systems with open boundary conditions, we
must build the doubled transition operator \riv, given by
\eqn\ei{
U_{a,b} \left( \theta \right) = T_{a,c} \left( \theta \right) K_{c,d}^{+}
\left( \theta \right) T_{d,b}^{-1} \left(- \theta \right)
}
with
\eqn\eii{
T_{a,c} \left( \theta \right)  =
L_{a,a_{1}} \otimes L_{a_{1},a_{2} }\otimes \cdots \otimes
 \L_{a_{N-1},c}.
 }
A graphic representation of the $L$ operator is given in fig. 1. Tensorial
product in the site spaces (vertical indices) and the usual
 product of matrices in the auxiliary space (horizontal indices) is understood
{}.
\bigskip
\centerline{\epsfxsize=8cm  \epsfbox{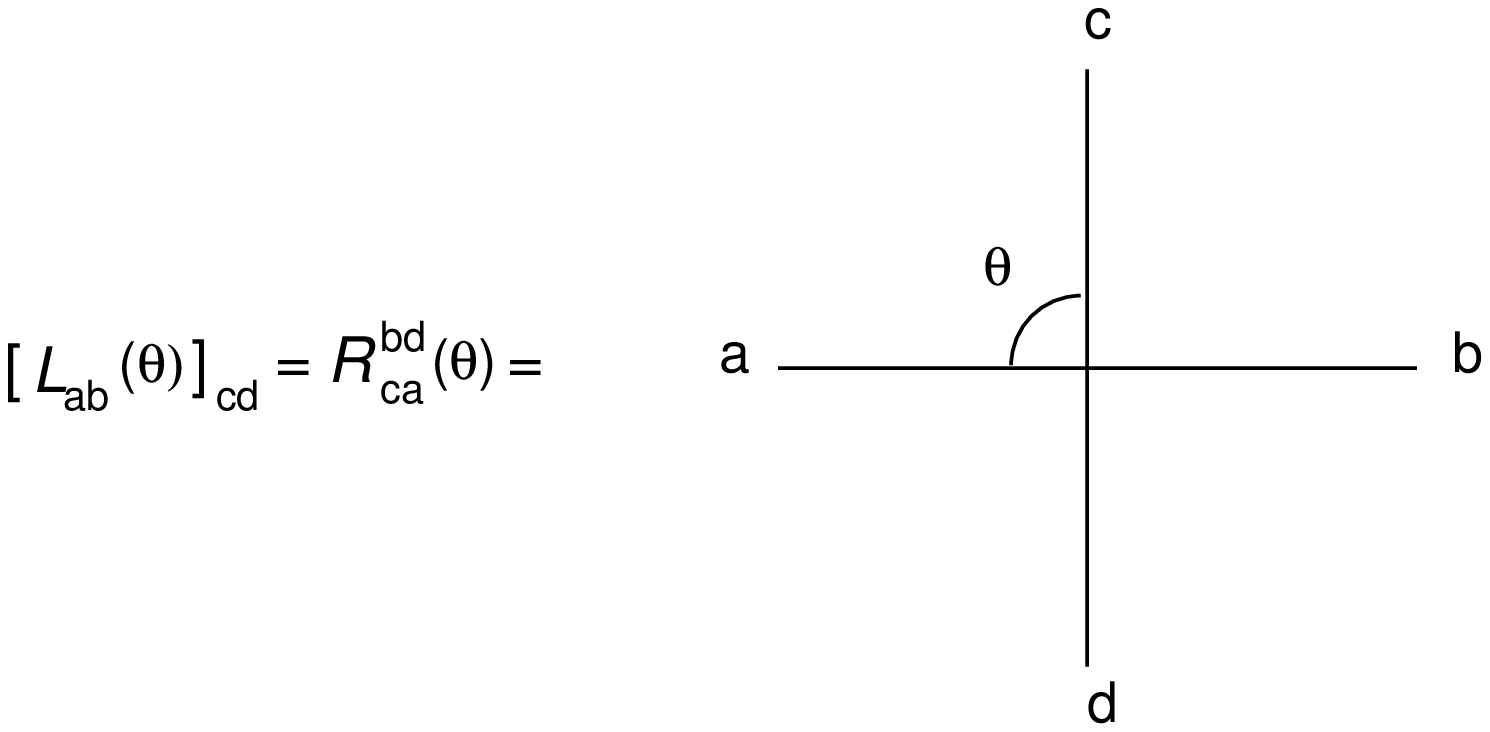}}
\centerline{Fig. 1}
\bigskip

The $K^{+} \left( \theta \right) $ matrix, which determines the boundary
condition
must verify the well known reflection condition expressed by
\
\eqn\eiv{\eqalign{
R(\theta-\theta') \left[  K^{+} (\theta) \otimes I\right] &R(\theta+\theta')
\left[  K^{+} (\theta')  \otimes I\right]   =\cr
&\left[  K^{+} (\theta ') \otimes I\right] R(\theta+\theta')
\left[  K^{+} (\theta ) \otimes I\right] R(\theta-\theta').\cr
}}
This relation is graphically expressed in fig. 2.
\bigskip
\centerline{\epsfxsize=8cm  \epsfbox{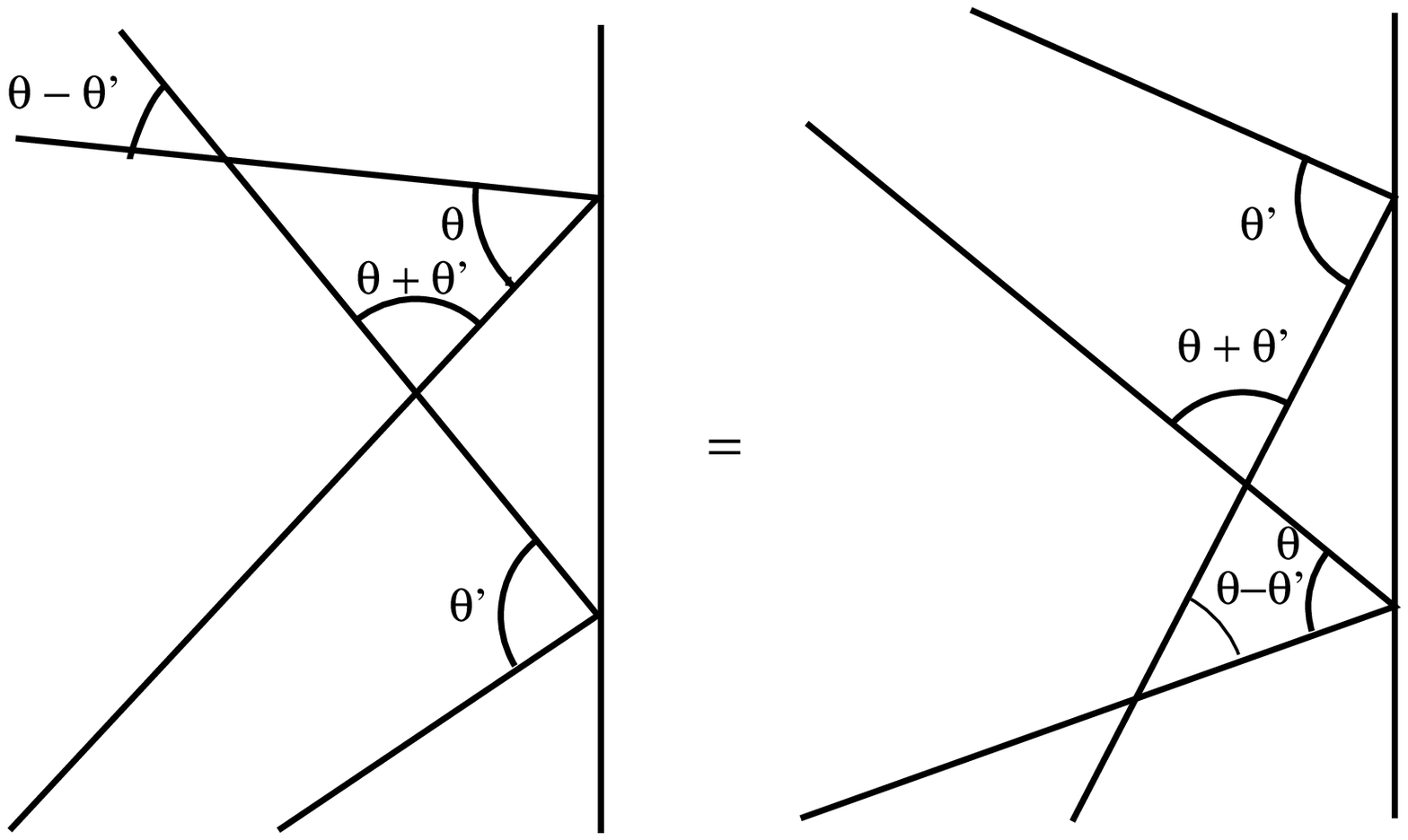}}
\centerline{Fig. 2}
\bigskip

Under this conditions, the operator
\eqn\ev{
t \left( \theta \right) =K_{a,b}^{-}(\theta) T_{a,c} \left( \theta \right)
K_{c,d}^{+} \left( \theta \right) T_{d,b}^{-1} \left(- \theta \right)
}
is a generator of constants of motion, since
\eqn\evi{
\left[ t\left( \theta \right) ,t\left(\theta' \right)\right] = 0,
}
and then, an open local hamiltonian, given by
\eqn\evii{
h={d {t\left( \theta \right)}\over d\theta}\bigg|_{\theta=0},
}
can be obtained.

If the system is invariant under parity, time reversal and crossing (case of
$A_{(1)}$), then the $K$ operator has the property
\eqn\eviii{
K^{+}(\theta)=K^{- t}(-\theta- \eta)
}
where $t$ means transposition and $\eta$ is the parameter of crossing property,
\eqn\eix{
R_{a,c}^{b,d} (\theta) R_{c,e}^{d,f} (-\theta - 2  \eta)
 \propto \delta _{a,e}  \delta _{b,f} .
}
On the other hand, if the system is invariant under $PT$ and not invariant
under crossing symmetry and if there is a matrix $M$ \rv--\rvii  such that
\eqn\ex{
R_{a,b} ^{c,d}(\theta) M_{b,e} R_{f,d}^{g,e} (-\theta - 2 \eta) M_{f,h}^{-1}
\propto \delta _{a,g}  \delta _{c,h} .
}
then
\eqn\exi{
K^{+}\left( \theta \right) = K^{- t}\left( -\theta-\eta \right) M^{- 1}.
}

General solutions for the boundary matrices have been found in the models
$XXX$, $XXZ$ and $XYZ$  \rix, and diagonal solutions for the models of type
$A_{(r)}$ with $r\geq 2$
diagonal solutions \rx.
In this paper we present non diagonal solutions for $su(3)$ and $su(4)$ models,
that we generalize afterwards to $su(n)$.

The matrix $R$ is given \ref\rxi{J. Abad and M. Rios, Integrable models
associated
to classical representations of $U_q\left( \widehat{sl(n)} \right)$. University
of Zaragoza preprint, DFTUZ 94/11.} by
\eqn\eni{
\eqalign{
R (\theta) =& \sinh{( {n \over 2} \theta +\gamma) }  \sum_{i=1}^{n}{
e_{i,i}\otimes e_{i,i} }+
\sinh{({n \over 2}  \theta)} \sum_{i,j=1 \atop i \neq j}^{n}{ e_{i,j}\otimes
e_{j,i} }    \cr &+
\sinh{(\gamma)}  \sum_{i,j=1 \atop i \neq j}^{n} {    exp{((j-i-{n \over 2}
\sign {(j-i)}) \theta)}
 e_{i,i}\otimes e_{j,j}   },\cr}
}
that in case $n=3$ becomes
\eqn\exii{
R(\theta)=
\left (\matrix{
a & 0 & 0 & 0 & 0 & 0 & 0 & 0 & 0 \cr
0 & d & 0 & b & 0 & 0 & 0 & 0 & 0 \cr
0 & 0 & c & 0 & 0 & 0 & b & 0 & 0 \cr
0 & b & 0 & c & 0 & 0 & 0 & 0 & 0 \cr
0 & 0 & 0 & 0 & a & 0 & 0 & 0 & 0 \cr
0 & 0 & 0 & 0 & 0 & d & 0 & b & 0 \cr
0 & 0 & b & 0 & 0 & 0 & d & 0 & 0 \cr
0 & 0 & 0 & 0 & 0 & b & 0 & c & 0 \cr
0 & 0 & 0 & 0 & 0 & 0 & 0 & 0 & a \cr
}\right ),
}
where
\eqna\exiii{
$$\eqalignno{
&a(\theta)=\sinh{({3 \over 2} \theta+\gamma)},&\exiii a\cr
&b(\theta)= \sinh{({3 \over 2} \theta)} ,&\exiii b \cr
&c(\theta)=\sinh{(\gamma)}  e^{{ \theta\over 2}},&\exiii c \cr
&d(\theta)=\sinh{(\gamma)} e^{{ -\theta\over 2}}.&\exiii d
\cr
}
$$}
This model is $PT$ invariant but not invariant under crossing symmetry, besides
this, it verifies the property
$\ex$
 with $M$ equal to the identity $I$ and
$\eta =\gamma$.

We look for solutions to equation \eiv\  with $K^{+} (\theta)$ of the form
\eqn\exiv{
K^{+} (\theta) =
\pmatrix{
x_{1}(\theta) & y_{1}(\theta) & z_{1}(\theta) \cr
x_{2}(\theta) & y_{2}(\theta) & z_{2}(\theta) \cr
x_{3}(\theta) & y_{3}(\theta) & z_{3}(\theta) \cr
}
}
and fulfilling the condition
\eqn\exv{
K^{+} (0) =I
}

As stated before, we are interested in non-diagonal solutions for
$K^{+}(\theta)$. The resulting set of equations has not solution with all
elements out of the diagonal in \exiv\ different from zero. Those equations,
instead, are satisfied when only one non-diagonal element and its symmetric are
allowed to be different from zero, for instance,
\eqn\exvi{
K^{+} (\theta) =
\pmatrix{
x_{1}(\theta) & 0 & z_{1}(\theta) \cr
0 & y_{2}(\theta) &0\cr
x_{3}(\theta) &0 & z_{3}(\theta) \cr
}
}

The solution that we find is
\eqna\exvii{
$$\eqalignno{
x_{1}(\theta) & = K_{a +} \sinh{(\epsilon_{+} -{3\over 2 }\theta)}&\exvii a \cr
z_{1}(\theta) & = K_{d +} e^{\theta \over 2}\sinh{(3  \theta)}&\exvii b \cr
y_{2}(\theta) & = K_{a +} e^{-\theta}\sinh{(\epsilon_{+} +{3\over 2} \theta)}+
       K_{b +} e^{\theta \over 2}\sinh{(3 \theta)}&\exvii c \cr
x_{3}(\theta) & = K_{c +} e^{\theta \over 2}\sinh{(3 \theta)}&\exvii d \cr
z_{3}(\theta) & = K_{a +} e^{\theta}\sinh{(\epsilon_{+} +{3\over 2}
\theta)}&\exvii e \cr
}$$
}
where the $K$'s must be verified the relations
\eqn\exviii{
K_{c+} K_{d+}=K_{b+} \left( K_{b+} +K_{a+} e^{- \epsilon_{+}}\right).
}
Of course, if we impose $ K_{b+}=0$, we obtain the diagonal solutions \rix,
\rx.

With this solutions, we can obtain through \exi\ the expression for the matrix
$K^{-}(\theta)$  and then the hamiltonian \evii. If we redefine the hamiltonian
by subtracting a constant times the identity, in order to eliminate constant
terms proportional to the length of the chain, it becomes
\eqn\exx{
\eqalign{
H^{su(3)}=& {1 \over 2} \trace\left( K^{-}(0) \right)
\sum_{n=1}^{N-1}{\left( J_9 (\lambda_n^3 \lambda_{n+1}^8 -\lambda_n^8
\lambda_{n+1}^3 )+\sum_{\alpha =1}^{8}{\left( J_\alpha \lambda_n^\alpha
\lambda_n^\alpha  \right)} \right)} +\cr
&+J_{1,3} \lambda_1^3 +J_{1,8} \lambda_1^8 +J_{1,4} \lambda_1^4 +
J_{1,5} \lambda_1^5  + \cr
&+ \trace\left( K^{-}(0) \right)
\left( J_{N,3} \lambda_N^3 +J_{N,8} \lambda_N^8 +J_{N,4} \lambda_N^4 +
J_{N,5} \lambda_N^5  \right) \cr
}
}
with
\eqna\exxi{
$$\eqalignno{
&J_1 =J_2 =J_4 =J_5 =J_6 =J_7 = 1  &\exxi a  \cr
&J_3 =J_8 =\cosh{\gamma }  &\exxi b \cr
& J_9={\sinh{(\gamma)} \over \sqrt{3} }  &\exxi c \cr
&J_{1,3}= { {e^{{-{ 3 \gamma} \over {2}}}\sinh{(3 \gamma)}}\over {3 \gamma}}
\left(  K_{b-} (1+e^{2 \epsilon_{-}+\gamma})+ {{K_{c-}{K_{d-}} \over {K_{b-}}}}
(e^{2 \gamma}+e^{2 \epsilon_{-}+\gamma})  \right)&\exxi d \cr
&J_{1,8}= { {e^{{-{ 3 \gamma} \over {2}}}\sinh{(3 \gamma)}}\over {\sqrt{3 }
\gamma}}
\left(  -K_{b-} (1+e^{2 \epsilon_{-}+\gamma})+ {{K_{c-}{K_{d-}} \over
{K_{b-}}}}
+e^{2 \epsilon_{-}+\gamma}  \right)&\exxi e \cr
&J_{1,4}={ { -2 e^{-{  \gamma \over 2}} \sinh{(3 \gamma)}}\over{
\sinh{\gamma}}}
\left( K_{d-} +K_{c-} \right)  &\exxi f  \cr
&J_{1,5}={ { - i 2 e^{-{  \gamma \over 2}} \sinh{(3 \gamma)}}\over{
\sinh{\gamma}}}
\left( K_{d-} -K_{c-} \right)  &\exxi g  \cr
&J_{N,3}={1 \over \sinh{\gamma}}   \left( K_{b+} ( e^{ 2\epsilon_{+}}-1 ) -
{{K_{c+} K_{d+} } \over K_{b+} } (2+e^{ 2\epsilon_{+}} )\right) &\exxi h  \cr
&J_{N,8}={\sqrt{3}  \over \sinh{\gamma}}   \left( K_{b+} ( e^{ 2\epsilon_{+}}+1
) -
{{K_{c+} K_{d+} } \over K_{b+} } e^{ 2\epsilon_{+}} \right) &\exxi i  \cr
&J_{N,4}={ 2\over \sinh{\gamma}} \left( K_{d+} +K_{c+} \right)  &\exxi j  \cr
&J_{N,5}={ {2 i}\over \sinh{\gamma}}\left( K_{d+} -K_{c+} \right)  &\exxi k
\cr
}$$
}
and
\eqn\exxi{
\eqalign{
\trace\left( K^{-}(0) \right)=& \left(  2\cosh{\gamma} \sinh{(\epsilon_{-}-{3
\gamma \over 2})}+\sinh{(\epsilon_{-}+{3 \gamma \over 2})}   \right)
\left( {K_{c-} K_{d-}  \over K_{b-} } -K_{b-} \right)  \cr
&- K_{b-} e^{-{\gamma \over 2}} \sinh{(3 \gamma)} .\cr
}}
In these expressions, we have removed the variables $K_{a+}$ and $K_{a-}$ by
using
\exviii\ and a similar expression with the subindices $(+)$ replaced by $(-)$.
So, the free parameters that we have in our solutions are
$$
\epsilon_{+}, \epsilon_{-}, K_{b+}, K_{b-}, K_{c+}, K_{c-}, K_{d+}, K_{d-}.
$$

The generalization to $su(n)$ follows from \exvii\ . The diagonal solution
\eqn\exxv{
K^{+}(\theta)=K^{+}_{i,i} \delta_{i,j}
}
similar to that obtained in ref. \rx , is
\eqn\exxvi{
\eqalign{
K^{+}_{j,j}(\theta)=&-\rho_{+} e^{\left( 2j+{n \over 2} \right) \theta}
\sinh{(\epsilon_{+}+
{n \over 2}\theta)}\qquad\hbox{for $1\leq j \leq l_{-}$} \cr
K^{+}_{l,l}(\theta)=&\rho_{+} e^{\left( 2l-{n \over 2} \right) \theta}
\sinh{(-\epsilon_{+}+
{n \over 2}\theta)}\qquad\hbox{for $l_{-}+1\leq l \leq n$} \cr
}
}
being $\rho$ and $\epsilon$ continuous parameters and $l_{-}$ a discrete
parameter
with integer values $l_{-}\leq n-1$.

The non-diagonal solutions are
\eqna\exxviii{
$$\eqalignno{
K^{+}_{1,1}(\theta)=&\rho_{a+} \sinh{(\epsilon_{+}-
{n \over 2}\theta)},  &\exxviii a \cr
K^{+}_{n,n}(\theta)=&\rho_{a+}  e^{(n-2)\theta}\sinh{(\epsilon_{+}+
{n \over 2}\theta)} ,&\exxviii b \cr
K^{+}_{j,j}(\theta)=&\rho_{a+}  e^{(2 j-n-2)\theta}\sinh{(\epsilon_{+}+
{n \over 2}\theta)} +
\rho_{b+}  e^{(2 j-2-{n \over 2})\theta}\sinh{(n \theta)},\hfill \cr
 &\qquad\qquad\qquad\qquad\qquad\qquad\qquad\qquad\qquad \hbox{$ j=2\cdots
n-1,$}
&\exxviii c \cr
K^{+}_{1,n}(\theta)=&\rho_{d+}  e^{({n \over 2}-1)\theta}\sinh{(n
\theta)},&\exxviii e \cr
K^{+}_{n,1}(\theta)=&\rho_{c+}  e^{({n \over 2}-1)\theta}\sinh{(n
\theta)},&\exxviii e \cr
}$$
}
the others components being zero. As before,
the $\rho$'s parameters must verify the relation
\eqn\exxvix{
\rho_{c+} \rho_{d+}= \rho_{b+}\left( \rho_{b+}+\rho_{a+} e^{-\epsilon_{+}
}\right).
}

The solution \exvii\ for $n=3$ is trivially obtained from \exxviii\ with the
$\rho$'s parameters substituted by  $K$'s ones.

As conclusions, we have shown a set of non-diagonal solutions for the boundary
matrices in an open $su(n)$ spin chain. Others solutions, with the non zero
elements out of the diagonal occupying others symmetric positions,  can be
obtained by means of similarity transformations.

In the analysis of the set of equations \eiv\ and the solutions  presented
here, a heavy use of  Mathematica computer program has been made.

{ \tenbf  Acknowledgements}

We thank to professor J. Sesma for the careful reading of the manuscript and
professor P. Kulish by useful comments. This work was partially supported by
the Direcci\'{o}n General de Investigaci\'{o}n Cient\'{\i}fica y T\'{e}cnica,
Grant No PB93-0302 and AEN94-0218

\listrefs
\end{document}